\documentclass{article}
\usepackage{amssymb}
\usepackage{amsmath}
\usepackage{amsfonts}
\usepackage{aip}

\input{tcilatex}

\begin{document}

\title{Magnetic-field-induced supercurrent enhancement in hybrid
superconductor/magnet metal structures }
\author{V.N. Krivoruchko \and Donetsk Physics \& Technology Institute NASU,
\and R. Luxemburg Str., 72, Donetsk-114, 83114 Ukraine}
\date{The Date }
\maketitle

\begin{abstract}
The dc transport properties of the (S/M)I(M/S) tunnel structure - proximity
coupled superconductor (S) and magnetic (M) layers separated by an insulator
(I) - in a parallel magnetic field have been investigated. We choose for the
M metal the one in which the effective magnetic interaction, whether it
arises from direct exchange interaction or due to configuration mixing,
aligns spins of the conducting electrons antiparallel to the localized spins
of magnetic ions. For tunnel structures under consideration, we predict that
there are the conditions when the destructive action of the internal and
applied magnetic fields on Cooper pairs is weakened and the increase of the\
applied magnetic field causes the field-induced enhancement of the\ tunnel
critical current. The experimental realization of the novel interesting
effect of the interplay between superconducting and magnetic orders is also
discussed.

\end{abstract}

With recent experimental observations of the $\pi $-phase state with the
critical current inversion in superconductor (S) - ferromagnet (F) hybrid
structures [1-3] and theoretical prediction of the supercurrent enhancement
in SF tunnel structures with very thin F layers [4-6], systems exhibiting a
nontrivial interplay between magnetism and superconductivity attract a lot
of attention. A common drawback of the FS systems is that, in ferromagnetic
metals, the exchange field acting on the spin of conducting electrons is in
general so large as to suppress superconductivity. Several options of how to
enhance the superconductivity of nanoengineered SF structures have recently
been discussed in the literature. In particular, magnetic-field-induced
superconductivity is predicted and observed\ [7] in
superconductor/lattice-of-magnetic-nanodots system due to the compensation
of the applied field between the dots by the stray field of the dipole array.

In the general case, when an external magnetic field is applied,
superconductivity is suppressed due to both orbital and spin pair breaking
effects. However, there are magnetic metals, such as (EuSn)Mo$_{6}$S$_{8}$
[8,9] or\ (MoMn)Ga$_{4}$ [10] where the applied magnetic field can induce
superconductivity. Several mechanisms that may enable superconductivity to
develop in such materials have been investigated in more or less detail (see
[11,12] and references therein). In the pseudoternary compounds, field
induced superconductivity is assumed to be due to the so-called
Jaccarino-Peter\ compensation effect [13]. It takes place in ferro- or
paramagnetic metals where, due to Hund coupling energy, the exchange
interaction, $J\mathbf{sS}$ , orients the spins $\mathbf{s}$ of the
conducting electrons antiparallel to the spins $\mathbf{S}$ of rare earth
magnetic ions. In such magnetic metals, the effective filed acting on the
spin of conducting electron is $\mathbf{H}+J<\mathbf{S}>$ with $J<0$. I.e.,
the exchange field \ $J<\mathbf{S}>$ can be reduced by the external magnetic
field $\mathbf{H}$ and the destructive action of both fields on the
conducting electrons can be weakened or even canceled. If, in addition,
these metals posses an attractive electron-electron interaction, it is
possible to induce bulk superconductivity by magnetic field.

In this report, we investigate a way to enhance the superconducting
properties of proximity coupled superconductor-magnetic (M) metal hybrid
structures by choosing the M metal with some specific properties. Namely, we
suppose that in the M film, due to Hund rules, the localized magnetic
moments of the ions, oriented by magnetic field, exert the effective
interaction, $H_{E}$, on spins of the conduction electrons. The latter,
whether it arises from the usual exchange interaction or due to
configuration mixing, is the antiferromagnetic type. In particular,\ such
material can be a thin layer of the pseudoternary compounds like (EuSn)Mo$%
_{6}$S$_{8}$ or dilute superconducting systes as Mo$_{77}$Ir$_{23-x}$Fe$_{x}$
[12], or some ferromagnetic intermetallic compounds. (While experimentally
the compensation effect was observed [11,12] for paramagnets, the
Jaccarino-Peter mechanism is applicable both to ferromagnetic and
paramagnetic metals, and both type of the orders will be assumed here.)
There are no specific requirements to the superconductor, so that it can be
any superconducting film proximity coupled with the M metal. We will
consider the layered S/M system under the effect of a parallel magnetic
field. It should be noted that the applied magnetic field is too weak to
induce the superconducting properties through the Jaccarino-Peter scenario,
if the M metal is the pseudoternary compound, i.e., we suppose that \textit{%
superconductivity of the M metal is due to proximity effect }. To be
definite, we calculate the dc critical current of the tunnel structure where
both electrodes are proximity coupled S/M bilayers in weak external magnetic
field. It will be demonstrated that in the region where the destructive
action of the fields is decreased, an increase of the\ magnetic field causes
the enhancement of the Josephson critical current.

The system we are interested in is the (S/M)I(M/S) tunnel structure of the
superconducting S/M bilayers separated by an insulating barrier (I) (see
fig.1). Let us assume that both films are very thin: i.e., $d_{S}\ll (\xi
_{S},\lambda _{S})$\ \textbf{, }$d_{M}\ll (\xi _{M},\lambda _{M})$. Here $%
\xi _{S(M)}$ is the superconducting coherence length of the S(M) layer; $%
\lambda _{S(M)}$\ is the London penetration depth of\ the S (M) layer. To
tackle the physics, we will suppose that the S and M metals are in good
electric contact and the transparency of the insulating layer is small
enough to neglect the effect of a tunnel current on the superconducting
state of the electrodes. Longitudinal dimension of the junction, $W$, is
supposed to be much less than the\ Josephson penetration depth, $W\ll
\lambda _{J}$ , so that a flux quantum can not be trapped by the junction.

As far as the thicknesses of the films are small, it is reasonable to assume
that magnetic field is homogeneous in the S/M bilayer. The conditions ensure
also that the orbital effects can be neglected. Also, in the limit $d_{S}\ll
\xi _{S}$\textbf{\ , }$d_{M}\ll \xi _{M}$, the influence of the M layer on
superconductivity in the S/M bilayer is not local\ and is equivalent to
inclusion of a homogeneous exchange field with a reduced value. Other
physical quantities characterizing the S metal in the S/M bilayer should be
modified, as well. Such an approach was recently discussed in\ [4,14] for
SFIFS structures, and, as was demonstrated, under these assumptions, a thin
S/F bilayer is equivalent to a superconducting ferromagnetic film with a
homogeneous superconducting order parameter and an effective exchange field.
Similarly, we can characterize the S/M bilayer by the effective values of
the superconducting order parameter $\Delta _{ef}$, the coupling constant $%
\gamma _{ef}$ and the exchange field $H_{Eef}$ described by the relations:%
\begin{eqnarray}
\Delta _{ef}/\Delta _{0} &=&\gamma _{ef}/\gamma =\nu _{S}d_{S}(\nu
_{S}d_{S}+\nu _{M}d_{M})^{-1}, \\
H_{Eef}/H_{E} &=&\nu _{M}d_{M}(\nu _{S}d_{S}+\nu _{M}d_{M})^{-1},
\end{eqnarray}%
where $\nu _{S}$ and $\nu _{M}$ are the densities of quasiparticles states
in the superconductor and magnetic metal, respectively; $\Delta _{0}=\Delta
(0,0)$ is the BCS value of the superconducting order parameter of the S
metal at T = 0 in the absence of the applied magnetic field, $\gamma $\ is
the coupling constant in the S metal. If the M metal is the pseudoternary
compound and can posses a nonzero electron-electron interaction, we will
neglect this interaction, so that the relations (1) remain valid in the case
as well.

The low transparency of the junction barrier allows to use the relation of
the standard tunnel theory [15]. According to this theory, the distribution
of the Josephson current density $j_{T}(x)$ flowing in z-direction through
the barrier takes the form $\ j_{T}(x)=j_{C}\sin \varphi (x)$ , where $%
\varphi (x)$ is the phase difference of the order parameter across the
barrier. In the case of a finite electrode thickness, the phase difference
of the order parameter is described by the well known equations [16]. The
Josephson current density maximum, $j_{C}$, is determined by the electrode
properties and here we focus on calculation of the $j_{C}$ .

Assuming that the exchange field $H_{Eef}$\ and the external magnetic $H$\
field act only on the spin of electrons, and in the conventional singlet
superconducting pairing, we can write the Gor'kov equations for the S/M
bilayer in the magnetic field in the form:\ 
\begin{align}
\lbrack i\varepsilon _{n}-\xi -(H_{Eef}-H\mathbf{)]}G_{\varepsilon
\upuparrows }+\Delta _{ef}F_{\epsilon \downarrow \uparrow }& =-1, \\
\lbrack i\varepsilon _{n}+\xi -(H_{Eef}-H)]F_{\varepsilon \downarrow
\uparrow }+\Delta _{ef}^{\ast }G_{\epsilon \upuparrows }& =0,
\end{align}%
where\ $\xi =\varepsilon (p)-\varepsilon _{F}$, $\varepsilon _{F}$ is the
Fermi energy, $\varepsilon (p)$ is the quasiparticle spectrum, \ $%
\varepsilon _{n}=\pi T(2n+1)$, $n=0,\pm 1,\pm 2,\pm 3,...$ are Matsubara
frequencies; $T$ is the temperature of the junction (we have taken the
system of units with $\hbar =\mu _{B}=k_{B}=1$); $G_{\varepsilon \upuparrows
}$ and $F_{\epsilon \downarrow \uparrow }$ are the normal and anomalous
Green functions, and $\uparrow ,\downarrow $ (or $\sigma =\pm 1$ in eqs.
(6), (7), below) is spin variable. The additional set of equations for $%
G_{\varepsilon \downdownarrows }$ and $F_{\epsilon \uparrow \downarrow }$\
can be readily written down from symmetry arguments. The equations are also
supplemented with the well known self-consistency equations for the pair
potential $\Delta _{ef}(T,|(H_{Eef}-H\mathbf{)|})$. In our case one can
easily obtain:%
\begin{equation}
\ln \left( \frac{\Delta _{0}}{\Delta }\right) =\tint_{0}^{\omega _{D}}\frac{%
dx}{\sqrt{x^{2}+\Delta ^{2}}}\{\frac{1}{\exp ([\sqrt{x^{2}+\Delta ^{2}}%
-(H_{Eef}-H\mathbf{)}]/T)+1}
\end{equation}%
\begin{equation*}
+\frac{1}{\exp ([\sqrt{x^{2}+\Delta ^{2}}+(H_{Eef}-H\mathbf{)}]/T)+1}\}.
\end{equation*}%
Here and below $\Delta \equiv \Delta _{ef}(T,|(H_{Eef}-H\mathbf{)}|)$; $%
\omega _{D}$ is the Debye frequency. If $H_{Eef}=H$ the formula (5) is
reduced to eq. (16.27) of Ref.17.

Following the Green's function formalism, the (S/M)I(M/S) tunnel junction
critical current can be written as follows:%
\begin{equation}
I_{C}=(2\pi T/eR_{N})\sum_{n,\sigma =\pm 1}f_{\epsilon \sigma
}(H_{Eef}-H)f_{\epsilon \sigma }^{+}(H_{Eef}-H),
\end{equation}%
where $R_{N}$ is the contact resistance in the normal state and $%
f_{\varepsilon \sigma }$ are averaged over energy $\xi $ the anomalous Green
functions $F_{\epsilon \sigma ,-\sigma }$ . One can easily find that:%
\begin{equation}
f_{\varepsilon \sigma }=\Delta ^{\ast }[(\varepsilon _{n}-i\sigma (H_{Eef}-H%
\mathbf{)})^{2}+\Delta ^{2}]^{-1/2}.
\end{equation}%
Using eqs. (6) and (7), after summation over spin index we find for the
reduced (i.e.$\ I_{C}eR_{N}\{4\pi T_{C}\Delta ^{2}\}^{-1}$) quantity 
\begin{equation}
j_{C}(T,H)=\frac{T}{T_{C}}\sum_{n>0}\frac{\varepsilon _{n}^{2}+\Delta
^{2}-(H_{Eef}-H)^{2}}{[\varepsilon _{n}^{2}+\Delta
^{2}-(H_{Eef}-H)^{2}]^{2}+4\varepsilon _{n}^{2}(H_{Eef}-H)^{2}}
\end{equation}%
The Josephson critical current of the junction, as function of the fields
and temperature, can be calculated using formula (8) and self-consistency
equation (5) [18]. In the general case, the dependence of the
superconducting order parameter on effective field can be complex enough due
to the possibility of transition to the nonhomogeneous
(Larkin-Ovchinnikov-Fulde-Ferrell) phase [19,20]. To keep the discussion
simple, we will not touch upon this scenario here, restricting the
consideration below to the region with the homogeneous superconducting
state. Even in this case at arbitrary temperatures the values of the $\Delta
_{ef}(T,|(H_{Eef}-H\mathbf{)}|)$ can be determined only numerically. The
phase diagram of a homogeneous superconducting state in the $H-T$ plane has
been obtained previously [14]. At finite temperatures, it is found that $%
\Delta (T,H)$ has a sudden drop from a finite value to zero at a threshold
of $H$, exhibiting a first-order phase transition from the superconducting
state to normal state. Using the results of Li \textit{et al}., from Eq. (5)
we take only one branch of solutions, corresponding to a stable homogeneous
superconducting state.

We are now able to analyze the critical current dependence on the fields
value and temperature. Figure 2 shows the results of numerical calculations
of the expression (8) for the Josephson critical current versus external
magnetic field for the case of low $T=0.05\Delta _{0}$\ and medium $%
T=0.2\Delta _{0}$ temperatures, and different values of the exchange field.
As is seen in fig.2, for some interval of the applied magnetic field the
enhancement of the dc Josephson current takes place in comparison with the
case of $\ H=0$. Note that, in the range of our formulas validity, the
larger the effective field $H_{Eef}$ is, the larger growth of the critical
current can be observed (compare the $j_{C}\ $curves for $H_{Eef}=0.25\Delta
_{0}$ and $H_{Eef}=0.6\Delta _{0}$\ at $H=0$ in fig. 2). This behavior is
also predicted by the expression (8). A sudden break off in $j_{C}(H)\ $%
dependences in the presence of $H$ results due to a first-order phase
transition from a superconducting state with finite $\Delta $ to a normal
state.

The magnetic-field enhancement of the critical current can be qualitatively
understood using the simple fact that the Cooper pairs consist of two
electrons with opposite spin directions. Pair--breaking effect due to
spin-polarized electrons is weakened, if the applied field increased
remaining $H\leq H_{Eef}$ , since the spin polarizations from the exchange
field of the magnetic ions and the applied field are of opposite signs and
reduce each other. On the other hand, the paramagnetic effect is again
increased, if the applied field increased for $H>H_{Eef}$. These
dependencies determine the (S/M)I(M/S) critical current behavior on the
field in the region $0\leq |H_{Eef}-H|<0.755\Delta $.

We emphasize that the scenario of the applied field enhancement of the
critical current differs from those studied before in [4,5,6,14] for SFIFS
tunnel structures. Note that the exchange field may increase $j_{C}$ of the
SFIFS junction for antiparallel mutual orientation of the layers
magnetization and only\textit{\ at low temperatures }$T\ll T_{C}$ [4,14]. In
our case the mechanism described above is valid for full temperature region
of the homogeneous superconducting state. To illustrate qualitative behavior
at finite temperatures, let us consider the case with $(\Delta
,|H_{Eef}-H|)\ll \pi T_{C}$ . Direct calculation of eq. (8) gives then for
the critical current 
\begin{equation}
j_{C}(H,\Delta )\sim \frac{\Delta ^{2}}{T\sqrt{\Delta ^{2}+(H_{Eef}-H)^{2}}}%
th(\frac{\sqrt{\Delta ^{2}+(H_{Eef}-H)^{2}}}{2T}),
\end{equation}%
If $\Delta \longrightarrow 0$ one obtains 
\begin{equation}
j_{C}(H)\sim \frac{\Delta ^{2}}{T^{2}}ch^{-1}(\frac{H_{Eef}-H}{2T}),
\end{equation}%
We also investigated [21] the case when only one electrode of a junction is
magnetic and the mechanism [4-6,14] definitely does not work - the SMIS
tunnel structures. The effect of magnetic-field-induced supercurrent
enhancement is predicted for such structures as well.

In conclusion, using specific properties of a magnetic material, we have
discussed a new way to enhance the superconductivity of superconductor -
magnetic metal hybrid structures by magnetic field. The idea is quite
straightforward: the magnetic metals are those where the effective magnetic
interaction, whether it arises from an exchange interaction or due to
configuration mixing,\ aligns the spins of the conducting electrons and the
magnetic ions in opposite direction. There are no specific requirements to
the superconductor proximity coupled with the magnetic metal. As predicted,
magnetic-field-induced enhancement of superconductivity of such hybrid
systems should be observed. To implement the idea, we consider the dc
Josephson effect for the (S/M)I(M/S) tunnel structure in parallel magnetic
field. Using approximate microscopic treatment of the S/M bilayer we have
demonstrated the effect of magnetic-field-induced supercurrent enhancement
in the tunnel structures. This striking behavior contrasts with the
suppression of the critical current by magnetic field. The existing large
variety of magnetic materials, the ternary compounds in particular, should
allow experimental realization of this interesting new effect of the
interplay between superconducting and magnetic orders.

\begin{center}
Figure captions
\end{center}

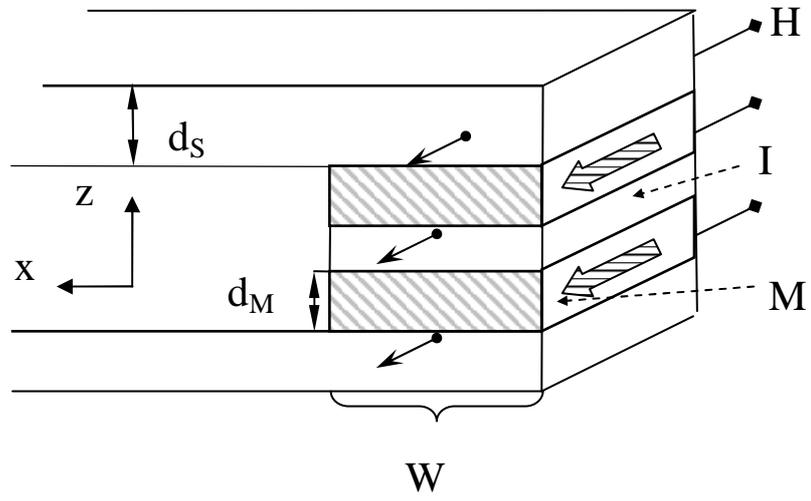
FIG. 1. (S/M)I(M/S) system in a parallel magnetic field. Here S is a
superconductor; M is a magnetic metal; I is an insulating barrier; W is
longitudinal dimension of the junction.

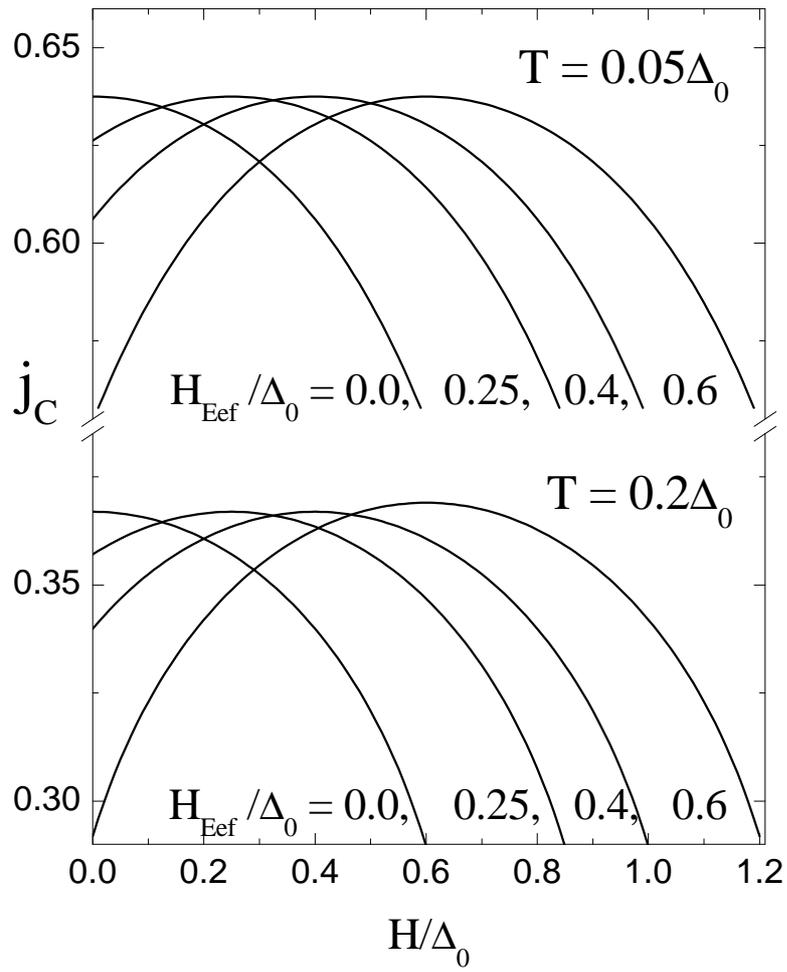
FIG. 2.\ Critical current of the SMIMS tunnel junction vs external magnetic
field for $T=0.05\Delta _{0}$\ , $T=0.2\Delta _{0}$ and different values of
the effective exchange field in the S/M bilayer: $H_{Eef}/\Delta _{0}=$ 0.0,
0.25, 0.4 and 0.6.


\begin{thebibliography}{99}
\bibitem{} As far as $H_{E}=J<\mathbf{S}>$ , a self-consistency equation
should be used for $H_{Eef}$, as well. However, we will suppose that $%
H_{Eef} $ , being much smaller than in isolated M film, is still larger than 
$\Delta _{ef}(T,|(H_{Eef}-H\mathbf{)}|)$ for full temperature region of the
homogeneous superconducting state. So, proceeding in the way to tackle the
new physics, we will ignore the temperature dependence of the $H_{Eef}$ .
\end{thebibliography}
\end{document}